\documentclass[3p,times,procedia]{elsarticle}
\usepackage{nupha_ecrc}
\newcommand{\rmd}{\mathrm{d}}
\newcommand{\rmi}{\mathrm{i}}
\newcommand{\rme}{\mathrm{e}}

\newcommand \beq{\begin{equation}}
\newcommand \eeq{\end{equation}}
\newcommand{\vp}{\vec{p}}

\volume{00}
\firstpage{1}
\journalname{Nuclear Physics A}
\runauth{}
\jid{nupha}

\jnltitlelogo{Nuclear Physics A}
\usepackage{amssymb}
\usepackage[figuresright]{rotating}
\begin{document}

\begin{frontmatter}

\dochead{XXVIth International Conference on Ultrarelativistic Nucleus-Nucleus Collisions\\ (Quark Matter 2017)}

\title{Phenomenological QCD equations of state \\ for neutron star mergers}

\author{Toru Kojo}
\address{Key Laboratory of Quark and Lepton Physics (MOE) and Institute of Particle Physics, Central China Normal University, Wuhan 430079, China}

\begin{abstract}
Thermal QCD equations of state at high baryon density are sensitive to the phase structure and the resulting excitation modes. The leading contribution at low temperature can be either $\sim p_F^2 T^2$ ($p_F$: quark Fermi momentum; $T$: temperature) for phases with gapless quarks, or $\sim T^4$ for phases with gapped quarks. In the latter the thermal pressure is dominated by collective modes. Starting with a schematic quark model developed for neutron star structure, we estimate the thermal contributions and zero point energy from the Nambu-Goldstone modes by building them upon the mean field background for the color-flavor-locked quark matter. Applying the phase shift representation for thermodynamic potentials, we include not only the bound state pairs but also resonating pairs. According to the Levinson's theorem, the high energy contributions tend to cancel the pole contributions to the thermodynamics, tempering the UV behaviors in the zero point energy. Our primary target in this talk is the domain with baryon density $n_B$ as large as $\sim 5-10n_0$ ($n_0\simeq 0.16\, {\rm fm}^{-3}$: nuclear saturation density), and the temperature $T$ of the order $\sim30-100\,{\rm MeV}$. The insights into this domain may be obtained through the future detection of gravitational waves from neutron star merging events. 
\end{abstract}


\end{frontmatter}

\section{Introduction}
\label{sec:intro}

The neutron stars (NSs) have QCD matter at almost zero temperature and at baryon density ($n_B$) reaching $\sim 5-10n_0$, where $n_0$ is the nuclear saturation density. In principle one can directly construct the QCD equation of state from the observed mass-radius ($M$-$R$) relations, and such attempts have been made within the current precision of the $M$-$R$ relations \cite{Ozel:2016oaf}, together with the constraints from nuclear experiments and many-body calculations \cite{Gandolfi:2013baa}. The recent trend indicates that QCD equations of state are soft at density $\lesssim 2n_0$ which leads to the radii $\lesssim 13\,{\rm km}$ \cite{Ozel:2016oaf}, but the matter must be very stiff at density $\gtrsim 5n_0$ to be consistent with the existence of two-solar mass NSs \cite{2m_1}. But in order to connect these soft and stiff equations of state, one must introduce the rapid change of pressure $P$ as a function of energy density $\varepsilon$ in the region between $\sim 2n_0$ and $\sim 5n_0$, which means that the (adiabatic) speed of sound $c_s^2 = \rmd P/\rmd \varepsilon$ must be very large; this would break the causality constraint $c_s^2 \le 1$. The situation becomes even severer if there is a strong first order phase transition which causes the significant softening, see \cite{Alford:2013aca} for systematic studies.

Extrapolating nuclear matter equations of state to very high density, it is often possible to satisfy all these constraints \cite{Akmal:1998cf}. But the validity of such procedures is doubtful beyond $n_B \sim 2n_0$; this is the region where many-body forces become comparable to two-body interactions, and hyperons appear and reduce the stiffness of matter, unless there are strong repulsions among hyperons and nucleons in not only two-body forces but also in three-body forces \cite{Lonardoni:2014bwa,Nishizaki:2001in}. As hadronic descriptions become inefficient, we need to gradually switch to the descriptions based on quarks.
The effective degrees of freedom should gradually change to avoid strong first order phase transitions, and this led several researchers to the descriptions based on the hadron-quark continuity which was originally proposed in \cite{Schafer:1998ef}, later discussed in the context of low temperature critical points \cite{Hatsuda:2006ps}, and recently utilized in the NS context \cite{Masuda:2012kf,Kojo:2014rca}. 

With the picture of hadron-quark continuity, a matter at $n_B \gtrsim 5n_0$ is described as a quark matter, and its equation of state is smoothly connected to a nuclear equation of state at $n_B \lesssim 2n_0$, with careful examination of the causality constraint for $n_B =2-5n_0$ \cite{Kojo:2014rca}. This procedure and the two-solar mass constraints put the severe restriction on the possible form of equations of state for $5n_0$. Within a schematic quark model of the Nambu-Jona-Lasinio (NJL) type, we examine the allowable range of the in-medium couplings of effective interactions and found that they must be as large as the vacuum couplings.
More details were reported in the previous conference (QM2015) \cite{Kojo:2015nzn}.

To get more detailed understanding of a matter, it is indispensable to study the excitations which reflect the phase structure and symmetry breaking. Such understanding is crucial in constructing thermal equations of state which are practically important for the descriptions of supernovae and NS merger events. In particular, during NS merger events the baryon density may reach $5-10n_0$ and the temperature is $\sim 30-100\,{\rm MeV}$ \cite{Sekiguchi}, and it may contain quark matter at finite temperature. In our quark model the matter at $n_B \gtrsim 5n_0$ becomes the color-flavor-locked (CFL) phase with quarks and gluons all gapped \cite{Alford:2007xm}, so that the leading thermal contributions come from the collective modes. In contrast to gapless quark matter with the large thermal pressure of $\sim p_F^2 T^2$ from quarks \cite{Kurkela:2016was,Masuda:2015wva}, the thermal pressure in CFL matter is dominated by the collective modes of $\sim T^4 \ll p_F^2 T^2$. For the use in gravitational simulations, we should also know the prefactor in addition to the parametric estimates, since the pressure variation of $\sim 10-30\, \%$ has deep impacts on NSs. Our estimates are based on the standard mean-field plus two-particle correlations. Through the computation of two-particle correlations, we also examine the potentially important zero-point energy.

\section{The collective modes}
\label{sec:NGmodes}

The lightest excitations in the CFL phase are Nambu-Goldstone (NG) modes. In the ideal chiral limit, the symmetry breaking pattern is $U(1)_B \times U(1)_Q \times SU(3)_L \times SU(3)_R \times SU(3)_C \rightarrow SU(3)_{C+L+R} \times U(1)_{ \tilde{Q} }$ ($B$: baryon number, $Q$: electric charge, $L,R$: chirality, $c$: color, and $\tilde{Q}$: modified electric charge) \cite{Alford:2007xm}. Through the symmetry breaking, the eight modes join the longitudinal modes of massive gluons, so 9 modes appear as NG modes. Furthermore, if the $U(1)_A$ symmetry is effectively restored, there is one extra NG mode which corresponds to $\eta'$. In QCD, the flavor symmetry is explicitly broken by the mass and electric charges of quarks, so NG modes except the one related to $U(1)_B$ are pseudo-NG modes which have nonzero masses. In addition, the imbalance in the size of the quark Fermi sea generates effective chemical potentials for $\pi^\pm$, $K^\pm$, and $K^0,\bar{K}^0$ \cite{Bedaque:2001je}. In particular the effective chemical potentials are comparable to the masses in the kaon channels, and it may lead to the kaon condensations in the CFL phase \cite{Kaplan:2001qk}.

These 10 NG modes have been studied by several groups mostly using the high density and/or weak coupling approximations \cite{Son:1999cm}. While these studies clarified the structure of NG modes, the direct applicability of these results for the NS physics is not obvious. At densities relevant to NSs, both the chiral and diquark condensates may coexist and we have to take into account the mixing of $q\bar{q}$ and $(qq)(\bar{q}\bar{q})$ fluctuations. There are not many such studies, and as far as we know, they were done before the two-solar mass constraints appeared \cite{Kleinhaus:2007ve,Yamamoto:2007ah}. For these reasons, we feel it necessary to update the computations by using quark models consistent with the current NS constraints \cite{Kojo:2016dhh,Song:2017dws}.

Following the standard computations, we calculate the NG modes in the RPA approximation using the vertices consistent with those used for the mean field calculations for quarks. For charged modes, we need to perform $3\times 3$ coupled channel calculations for particle-antiparticle, particle-particle, and antiparticle-antiparticle channels. For the neutral modes, we must compute $9\times 9$-coupled channels for $\pi^0,\eta,\eta'$.

\begin{figure}[tbp]
   \centering
   \vspace{-2cm}
   \includegraphics[width = 0.7\textwidth]{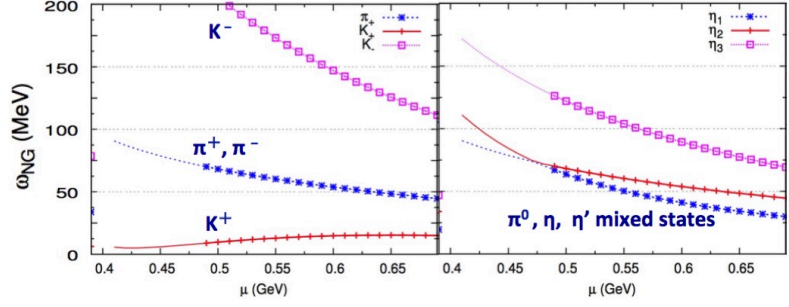} 
      \vspace{-2.2cm}
   \caption{The spectra of NG modes as functions of quark chemical potential $\mu_q$.}
        \vspace{-0.3cm}
   \label{fig:NGmodes}
\end{figure}

A brief summary of the results is the following \cite{Kojo:2016dhh} (Fig.\ref{fig:NGmodes}): (i) the NG modes except $K^+$ and $K^0$ have the masses in the range of $50-200\,{\rm MeV}$. Their masses decrease as the density increases and their structure is increasingly dominated by diquark components; (ii) the overall masses increase as we increase the coupling constants. Especially the light kaons are as heavy as $5-20\,{\rm MeV}$ and do not condense at strong coupling. In our setup for NSs, we did not find the kaon condensation at $n_B \gtrsim 5n_0$. But this issue is rather subtle and quantitive, and more detailed studies are called for.

%
\section{The phase shift representation for thermodynamic potentials}
\label{sec:phase}


For computations of thermodynamics, we adopt a unified description which includes not only the NG modes, but also the resonating pairs beyond the continuum thresholds. Such description is possible in the form of the phase shift representation of the thermodynamics \cite{Beth:1937zz,Dashen:1969ep,Zhuang:1994dw}. For a channel $X$,
\beq
\Delta \Omega_X =  \int \! \frac{\, \rmd\vp\,}{\, (2\pi)^3 \,} 
\int_{-\infty}^\infty \frac{\, \rmd \omega \,}{2\pi} \left[ \frac{\, |\omega| \,}{2} + T \ln \left(1+\rme^{ -|\omega|/T } \right) \right] \frac{\, \rmd \varphi_X(\omega,\vp) \,}{\rmd \omega} \,,
\eeq
where the all information in the channel $X$ (including chemical potentials) is encoded in the phase shift $\varphi_X$, the phase of the {\it connected} two-particle Green's function, $G_X  = |G_X| \rme^{\rmi \varphi_X}$. If $\delta \varphi_X/\rmd \omega = \pi \delta(\omega-m_R)$, then it reproduces the ideal gas expression for a meson with the mass $m_R$. The expression contains the thermal contributions as well as the zero point energy, so the corrections exist even at zero temperature.

It should be kept in mind that the expression already contains the subtraction of disconnected or free mean field propagator from the full propagator, so {\it the derivative of the phase shift can be negative}. In fact, the Levinson's theorem, which was extended into the more general form in \cite{Dashen:1969ep}, says that 
\beq
\varphi_X (\omega=\infty) = \varphi_X(\omega=0) \,.
\eeq
This constraint means that whenever $\varphi_X$ increases from zero to some positive value, it must come back to zero at high energy, accompanying the negative derivative. For instance, at the location of bound states, the phase shift jumps by $\pi$, and after scattering channels open, it gradually relaxes to zero as the energy increases. The origin of this constraint is the conservation of the phase space available for the states with respect to the interactions; the deformation of spectra at low energy must be compensated at high energy.

The Levinson's constraint has important impacts on the zero point energy which looks badly divergent as $\sim \Lambda_{ {\rm UV} }^4$ if we 
throw away the negative contributions at high energy. Including the deformation of the spectra from low to high energy, the low and high energy contributions tend to cancel, so at given $\vp$, the integral over $\omega$ becomes finite. As a result, even for non-renormalized model such as the NJL type, the UV divergence is reduced from $\sim \Lambda_{ {\rm UV} }^4$ to $\sim \Lambda_{ {\rm UV} }^3$. Then, the final expression is given by subtracting the zero point energy at $\mu=T=0$, and the $\sim \Lambda_{ {\rm UV} }^3$ terms cancel. What we found in the end is that the zero point energy does not show sizable cutoff dependence at least for the variation of the momentum cutoff $|\vp|_{ {\rm max} } = 600-1200\,{\rm MeV}$. It seems that neither $\Lambda_{ {\rm UV} }^2$ nor $\Lambda_{ {\rm UV} }$ survive; the cancellation is much better than we expected.

\begin{figure}[tbp]
   \vspace{-2cm}
  \hspace{2.0cm}
   \includegraphics[width = 0.7\textwidth]{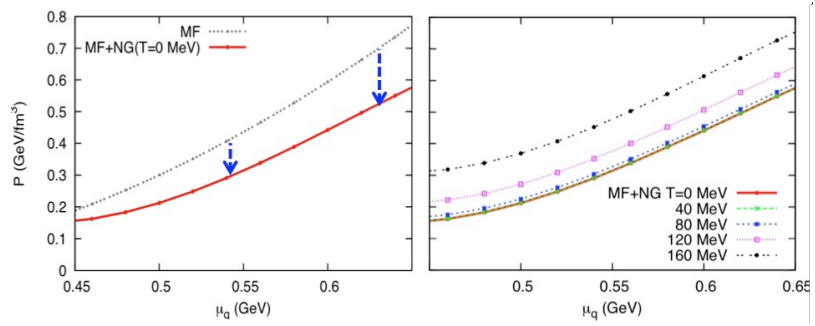} 
      \vspace{-2.cm}
   \caption{(Left) The mean field pressure plus the zero point energy. (Right) The thermal corrections from collective modes to the pressure.}
   \label{fig:zero_thermal}
           \vspace{-0.2cm}
\end{figure}

Including the pseudo-scalar channels for NG modes, the zero point energy reduces the mean field pressure at given chemical potential by $10-20\%$ (Fig.\ref{fig:zero_thermal}, left). This looks small compared to the mean field contribution, and our descriptions based on the mean field plus fluctuations look logically consistent. On the other hand, the modification requires us to slightly re-adjust model parameters for the NS structure calculations. 
As for thermal contributions, up to temperature $T\sim 40\,{\rm MeV}$ thermal corrections from the QCD sector are negligible (Fig.\ref{fig:zero_thermal}, right). In the CFL region, lepton densities are not high so those contributions are also of $\sim T^4$, with no significant enhancement due to the Fermi sea.

The present calculations are just tentative version of full calculations, and there are a lot of rooms for the improvements. Our studies are not very systematic yet; we need more studies of two particle correlations not only for the NG channels but also for the other channels, since for the zero point energy the degrees of freedom from low to high energy all contribute. These corrections will be presented elsewhere.

\bibliographystyle{elsarticle-num}
\bibliography{<your-bib-database>}

The author thanks the organizers for this enjoyable workshop. He also acknowledges G. Baym, Ho-Ung Yee, and R. Pisarski for discussions about the phase shift representation of thermodynamics, and Y. Sekiguchi for the explanations about gravitational waves from neutron star mergers. This work is supported by NSFC grant 11650110435.



\end{document}